# The time spectra in the neutron radiative decay experiment

Khafizov R.U. [a], Kolesnikov I.A [a]., Nikolenko M.V.[a], Tarnovitsky S.A.[a], Tolokonnikov S.V. [a], Torokhov V.D. [a], Trifonov G.M.[a], Solovei V.A.[a], Kolkhidashvili M.R. [a], Konorov I.V. [b]

[a] NRC «Kurchatov Institute», Russia
[b] Technical University of Munich, Munich, Germany

Annotation

To measure the main characteristics of radiative neutron decay, namely its relative intensity BR (branching ratio), it is necessary to measure the spectra of double coincidences between beta-electron and proton as well as the spectra of triple coincidences of electron, proton and radiative gamma-quantum. Analysis of double coincidences spectra requires one to distinguish events of ordinary neutron beta decay from the background; analysis of triple coincidences relies on distinguishing radiative neutron decay from background events. As demonstrated in our first experiment [1], these spectra presented a heterogeneous background that included response peaks related to the registration of electrons and protons by our electronic detection system. The NIST experimental group (emiT group) observed an analogous pattern on the spectrum of double coincidences [2]. The current report is dedicated to the analysis of this heterogeneous background. In particular, this report demonstrates that the use of response function methodology allows to clearly identify radiative neutron decay events and to distinguish them from the background. This methodology enabled us to become the first team to measure the relative intensity of radiative neutron decay B.R.= $(3.2\pm1.6)\cdot 10^{-3}$ (where C.L.=99.7% and gamma quanta energy exceeds 35 kev) [1]. In addition, the review emphasizes that the background events on the spectrum of double coincidences are caused by ion registration, and demonstrates that one cannot ignore the ionic background, which is why experiment [3, 4] registered the ions and not recoil protons.

**Introduction**

Presently, the characteristics of the ordinary decay mode are measured with precision of tenths of a percentage point. Under these circumstances, the experimental data obtained by different groups of experimentalists can be reconciled only by taking into account the corrections calculated within the framework of the standard theory of electroweak interactions. This means that the experimental research of the ordinary mode of neutron decay have exhausted their usefulness for testing the standard model. To test the theory of electroweak interaction independently, it is necessary to move from the research of the ordinary branch of decay to the next step, namely, to the experimental research of the radiative decay branch.

The radiative decay branch of elementary particles, where an additional particle, a radiative gamma quantum is formed along with the regular decay products, has been discovered for practically all elementary particles. This has been facilitated by the fact that among the rare decay branches the radiative branch is the most intensive, as its value is proportional to the fine structure constant α and is only several percent of the intensity of the regular decay mode (in other words, the relative intensity B.R. of the radiative decay branch has the value of several hundredths of a unit.)

However, for the neutron this decay branch had not been discovered until recently and considered theoretically only [1-4]. Our first attempt to register the radiative neutron decay events was made on intensive cold neutron beam at ILL [5]. But our experiment conducted in

2005 at the FRMII reactor of Munich Technical University became the first experiment to observe this elementary process [6]. We initially identified the events of radiative neutron decay by the triple coincidence, when along with the two regular particles, beta electron and recoil proton, we registered an additional particle, the radiative gamma quantum

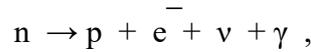

$$n \rightarrow p + e^- + \bar{\nu} + \gamma ,$$

and so could measure the relative intensity of the radiative branch of neutron decay B.R.= $(3.2\pm1.6)\cdot10^{-3}$ ( with C.L.=99.7% and gamma quanta energy over 35 kev; before this experiment we had measured only the upper limit on B.R. at ILL [5] ).

The main characteristic of any rare mode of elementary particle decay is its relative intensity, branching ratio (BR). By definition, BR is equal to the ratio between the intensity of the rare decay mode and the intensity of the ordinary mode. In the case of neutron, this intensity ratio can be reduced to the ratio between the number of triple coincidences between the registration of beta-electrons, radiative gamma-quantum and the delayed proton $N_T$ to the number of double coincidences between the registration of the ordinary decay products, beta electron and recoil proton $N_D$:

$$BR = I(\text{radiative decay}) / I(\text{ordinary decay}) = N(e,p,\gamma) / N(e,p) = N_T / N_D$$

These two values can be determined only from the analysis of double and triple coincidences spectra, which form corresponding peaks. Identifying these peaks and distinguishing them from the significant background is the central problem in the methodology of BR measurements for neutron radiative decay.

Further, this experimental BR value needs to be compared with the theoretical value, estimated within the framework of the electroweak model. Any difference between these two values would mean that we are observing a deviation from the electroweak interaction theory.

Our group calculated the neutron radiative spectrum in the framework of standard electroweak theory in the following papers [1-4]. The calculated branching ratio for this decay mode as a function of the gamma energy threshold was published in these papers. BR value for the energy region over 35 keV was calculated to be about $2.1\cdot10^{-3}$.

It follows that to measure the main characteristic of radiative neutron decay it is necessary to obtain and analyze the spectra of double and triple coincidences. So, it is necessary to consider the main particularity of these spectra - heterogeneous background.

Let us consider the reason behind the response peaks and show that their occurrence resulted from the way the detecting system works rather than from real physical events. We'll start from evaluating the simplest, ideal case where the detecting system emits rectangular impulses with front duration of zero, and the electron system uses a discriminator with a constant threshold. In this case the electron system creates a time-pickoff signal when the forward vertical front of the rectangular impulse arrives, and it gets registered in the corresponding time channel in the time spectrum. Fig.1 shows the rectangular impulse of duration $\tau_p$ with a flat line and under it there is a response peak which in this case is located in the only master channel that opens the time window with time channels of duration $\tau_c$ (in our case, the duration of the time channel equaled 25ns, and the window opened both forward and backward up to 100 channels). If the duration of the pulse exceeds the value of the time channel in several times, the master time channel that opens forward and backward time windows is distinct from the other channels. Fig.1 presents a case where the master channel opens the time window at a moment after the impulse has already arrived. In this case, the master channel appears to break the pulse into two parts: one part gets registered in the window opened backward, and the other part of the pulse will be registered in the master channel. It is obvious that in this case the channel is open not for the duration of time channel $\tau_c$, but rather for that value plus the duration of the pulse itself $\tau_p$, and then the channel would collect more pulses than the others. This, the time spectrum shows a response peak located in the only master channel (see the lower part of Fig 1), and this peak is filled with background pulses, with its height $H_R$ exceeding the background height $H_B$ by $(1+\tau_p/\tau_c)$ times.

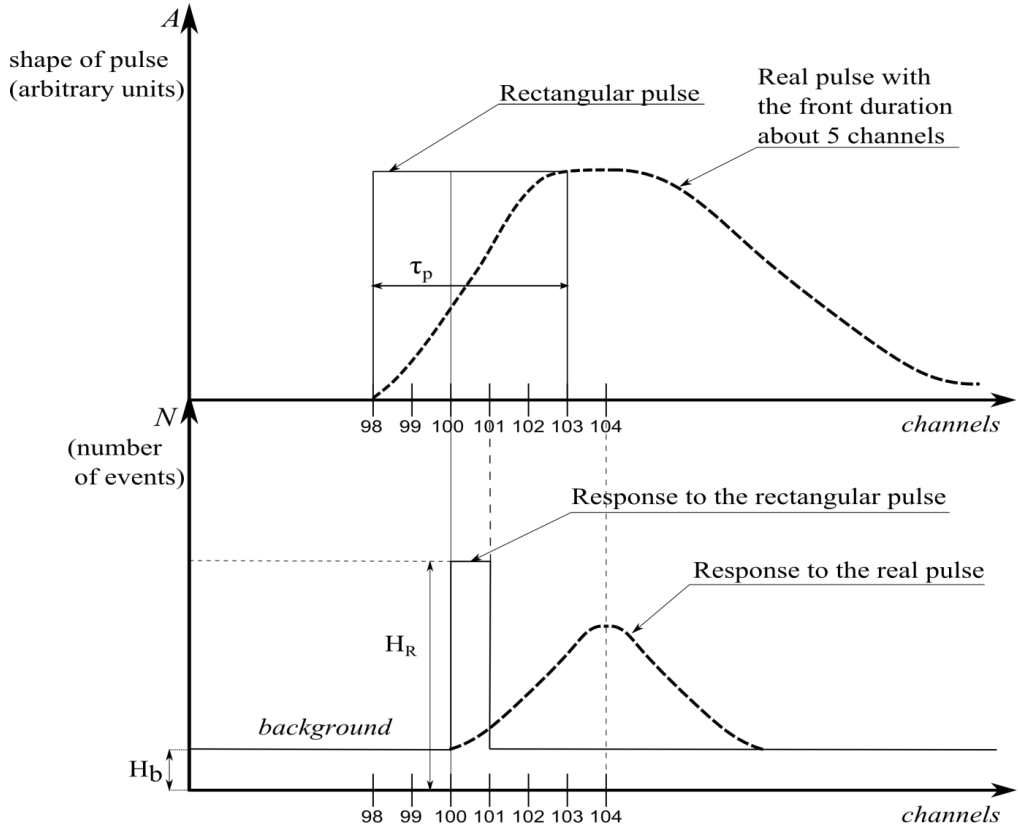

**Fig.1.** The top graph presents the shape of the impulses. The flat line shows the rectangular impulse with a zero front duration, while the dotted line represents the real front duration. The lower graph presents the form of the response peaks in time spectrum: the flat line shows the ideal case of rectangular impulses and the dotted line shows the real impulses.

But in our case the gamma pulse had a front of duration d that exceeded the duration of the time channel of 25 ns by several times, which makes it impossible to call it rectangular. Besides, instead of a discriminator with a constant threshold, we used constant fraction discriminator (CFD). This discriminator uses a line of delay that gives signal A(t-d) and a usual resistance divider that gives signal f•A(t). It compares the signal A(t-d) delayed by time d with the part of signal f•A(t). When the two signals are equal, the discriminator creates the time-pickoff signal, which falls into the corresponding time channel on the spectrum. In other words, this is the moment in which the below equation is solved:

$$f \cdot A(t) - A(t-d) = 0$$

It is obvious that the solution of this equation depends on the value of fraction f and the time of delay d. In our case we chose the delay time that would be equal to the duration of the front, so the solution of this equation would not coincide with the beginning of the pulse but rather is shifted relative to its beginning by a value comparable to the duration of the pulse front d. These circumstances lead to the fact the response peak is located not just in the master channel but is spread across several channels because of its width, with its maximum shifted by value comparable to the duration of the pulse front (see Fig 1, dotted line).

**Analysis and comparison of double coincidences spectra**

Here let's pause to analyze the spectra of double coincidences between beta-electron and recoil proton, and compare our spectrum with the results obtained by other authors. We have published the diagram of our experimental equipment in the past [5-6, 8]. Here we will simply note that in its sizes our equipment is comparable to the equipment used by the two other groups and the distance between the observed decay zone and the proton detector in our equipment is about 0.5m. The accelerating potential of the electric field is also approximately the same in all three equipment sets, so all three experiments should lead to similarly forms of double and triple coincidences spectra.

Fig.2 demonstrates the summary statistics on double e-p coincidences (coincidences of electron with delayed proton). Fig.2 clearly shows two major peaks: one peak with a maximum in channels 99-100, which is the peak of zero or prompt coincidences [6, 8]. The position of this peak marks the zero time count, namely the time when the electron detector registered the electron. This peak is not physics-related in its nature. Instead, it is a reaction of the detectors and the electronic system to the registration of the beta electron. It is namely the pulse from the electron channel that opens the time windows on spectra Fig.2 for 2.5 µs forward and backwards. The next peak visible on Fig. 2 has a maximum in channel 120 and is the peak of e-p coincidences of beta-electron with delayed proton.

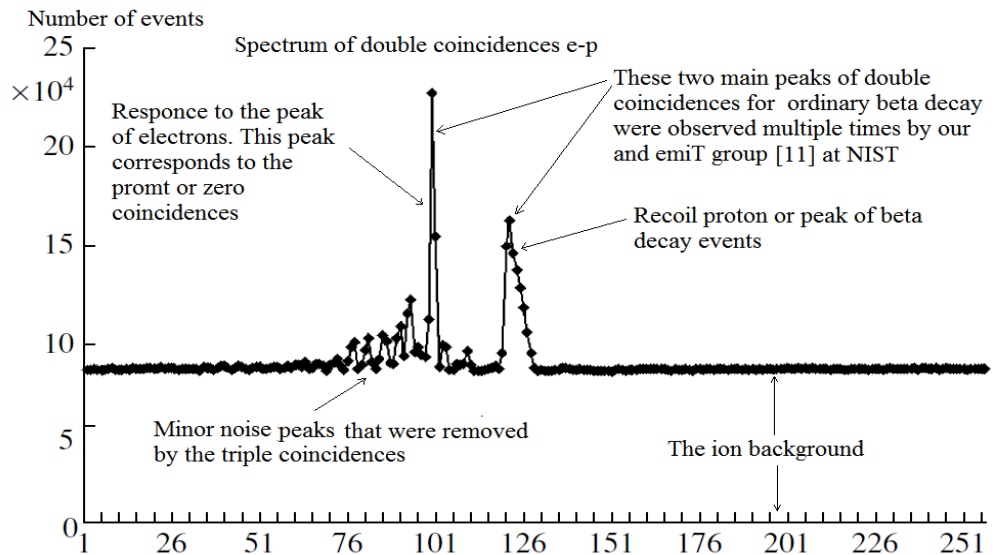

**Fig.2**. Timing spectrum for e-p coincidences. Each channel corresponds to 25 ns. The peak at channel 99-100 corresponds to the prompt (or zero) coincidences. The coincidences between the decay electrons and delayed recoil protons (e-p coincidences) are contained in the large peak centered at channel 120.

An analogous situation was observed in experiments on the measurement of the correlation coefficients by two independent groups at ILL [10] and emiT group at NIST [11], and it was also mentioned at [12]. We would especially like to emphasize the correspondence of our spectrum of double coincidences with an analogous spectrum from the result obtained by the emiT group from NIST [11]. On Fig. 3 we present their spectrum and diagram for the registration of the beta electron and the recoil proton. A comparison of our results with the results of the emiT group shows their unquestionable similarity. Moreover, the position of the second proton peak in Fig. 3 (emiT group), like in Fig. 2 (our result), corresponds well to the simple estimate obtained by dividing the length of a proton trajectory by its average speed.

Here we will also note the presence of a significant homogenous ionic background in Fig.2 and Fig.3. However, in both cases this background allows to easily distinguish the neutron decay peak. As we will shortly demonstrate, this ionic background will play a dominant role in the presence of a strong magnetic field and it will become impossible to distinguish events of ordinary neutron decay against it.

Following Avogadro's law, even in the case of a very deep vacuum under pressure of $10^{-6}$ – $10^{-8}$ mbar air molecule concentration remains very high. In fact, it is sufficient for beta-electrons produced in neutron decay to create a significantly high ionic background. Here one must note that the probability of ion creation along the trajectory of beta-electron in inverse proportion to the average distance between neighboring ions, i.e. proportional not to the molecule concentration but to the cubic root of this value. From this observation it follows that the value of the ionic background does not significantly depend on the vacuum conditions inside the experimental chamber. In our case, pressure was two orders of magnitude greater than the pressure in the emiT experiment. However, we observed an ionic background of only 4-5 times their background. This estimate is confirmed when one compares the spectra on Fig. 2 and Fig.3. Our spectrum, presented on Fig.2, has a 1:1 ratio of the value of e-p coincidences peak and the value of the background. The emiT group (Fig.3) spectrum has a ratio of 4:1 – 5:1, i.e. only 4-5 times our number, that is equal to the cubic root of pressure ratio in both teams' work.

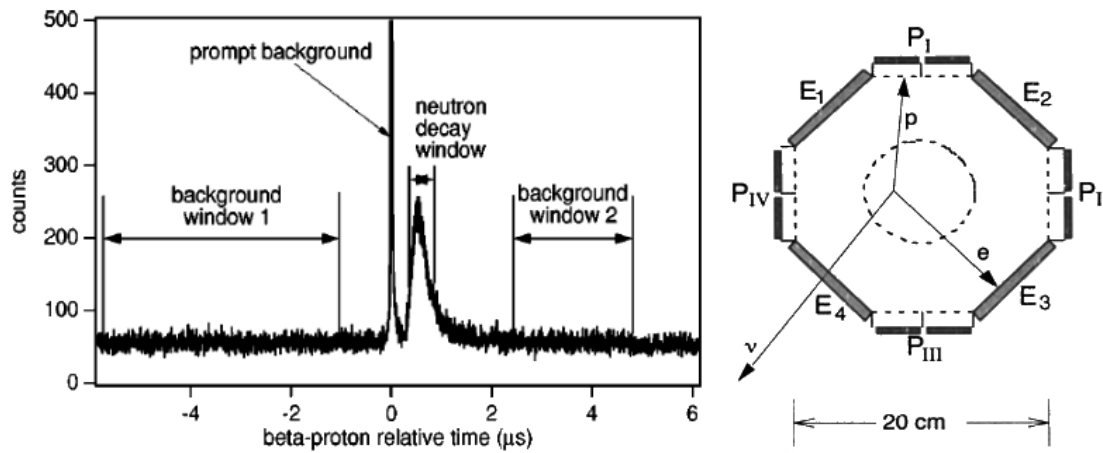

**Fig.3.** Spectrum of double electron-proton coincidences obtained by emiT Group [11] with two peaks and ion background value comparable to the neutron decay peak; emiT group scheme for registering beta electron and recoil proton.

Fig.2 shows that the total number of events in e-p coincidences peak in our experiment equals $N_D = 3.75 \cdot 10^5$. This value exceeds the value we obtained in our previous experiment conducted on beam PF1 at ILL by two orders of magnitude. It was precisely because of the low statistics volume that we could not identify the events of radiative neutron decay in that experiment and instead defined only the upper B.R. limit [5]. It is very important to note that the peak of double coincidences between electron and the delayed proton is observed against a non-homogenous background (see Fig.2 and Fig.3): besides the homogenous ionic background, which has a value comparable to the value of the e-p coincidences peak, there is an obvious peak in channels 99-100. In essence, this peak is a response peak to the time spectrum of electron registration, which contains just one peak in channels 99-100, signifying the time when the electron detector registered the electron. We will shortly see that the radiative peak of triple coincidences appears against a non-homogenous background with not one, but two response peaks.

The remaining peaks on Fig.2 are small, with just seven peaks distinct from the statistical fluctuations. These occurred because of the noise in the electric circuits of the FRMII neutron guide hall. There are no other physics-related reasons for their occurrence. These peaks appeared and disappeared depending on the time of day, reaching their maxima during the work day and disappearing over the weekends. Such behavior was observed throughout the experiment as we collected statistics. Since the nature of these seven small peaks is in no way related to radiative and ordinary decay, we did not emphasize them in our article.

The comparison conducted demonstrates that the spectra of double coincidences obtained in our experiment completely correspond with the results obtained by the emiT group. Now we will compare these two spectra with the spectrum of double coincidences obtained by the third group. Unfortunately, the authors did not publish the spectrum of double coincidences in their original article [7], instead it was only published this year in paper [13]. Fig 4 displays the spectrum of double coincidences and the diagram of their experimental equipment.

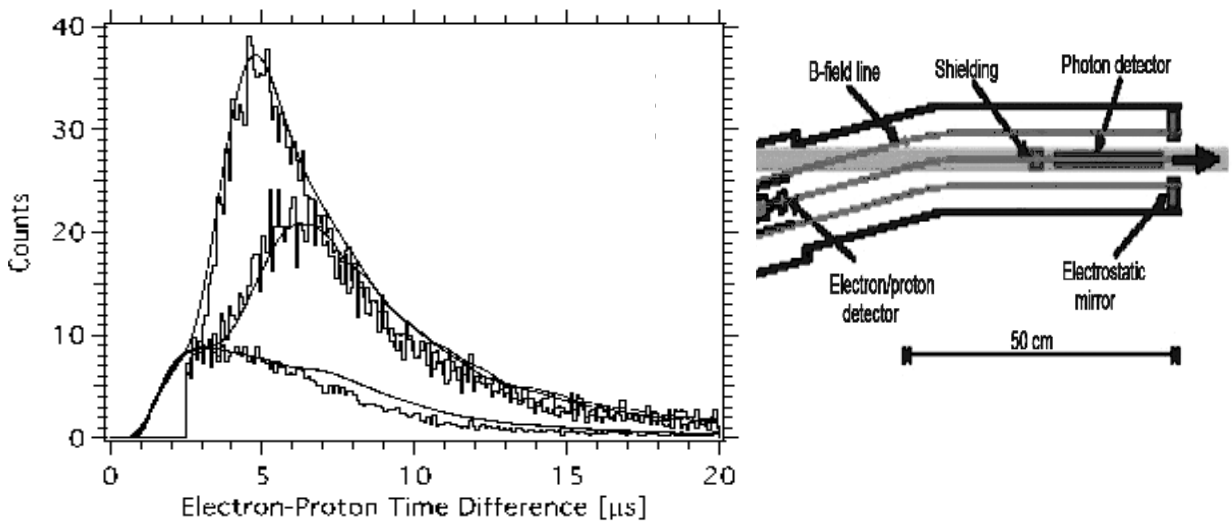

**Fig.4.** Equipment diagram and the single peak of "electron-proton" coincidences, published in [13]. The lower curve corresponds to 0 volts, the middle curve corresponds to 300 volts and the highest curve corresponds to 500 volts in an electrostatic mirror. The location of the peak's maximum and its significant width differ from our and the emiT results subsequently by one and two orders of magnitude. The location and the width of the peak also deviate by one and two orders of magnitude from the elementary estimates of delay times (see below).

The significant deviation obtained is explained by the fact that the peak in the NIST experiment consists not of beta-decay protons, but rather of ions. The density of gas molecules inside the equipment is proportional to pressure and according to the Avogadro's

Law is at the order of $10^7$ mol/cm$^3$ even at the pressure of $10^{-8}$–$10^{-9}$ mbar. This is a very significant number, which quite enough for creation the large ionic background in the presence of ionizing radiation. The energy of beta-electrons significantly exceeds the energy of ionization. Besides, the probability of ion creation by electrons is inverse proportional not to volume taken up by one molecule but to the average distance between molecules. It is precisely due to this reason that the ionic background falls proportionally to the cubic root of the pressure and not proportionally to pressure. In the emiT group experiment the pressure was the same as in the NIST experiment, i. so, the ionic background should be the same too. The light ions, together with the beta protons, should have a delay time comparable to 1 µs. The pulses from these particles are simply not visible in the spectrum due to the NIST group's use of combined electron-proton detector (see Figure 5 with the shape of electron and ion pulses). The maximum of the "proton" peak in the NIST experiment, according to the delay times estimations (delay time is proportional to square root of ion mass), falls exactly to the air ions 4-6 µs.

Fig.5 presents the pulse forms on the electron-proton detector. As was pointed out above, the significantly delayed pulses of low amplitude correspond to ion pulses, and the pulses from protons are simply invisible due to a presence of a wide electron pulse of high amplitude. Namely this fact explains the dead zone around zero on the spectrum of electron-ion coincidences on Fig.4.

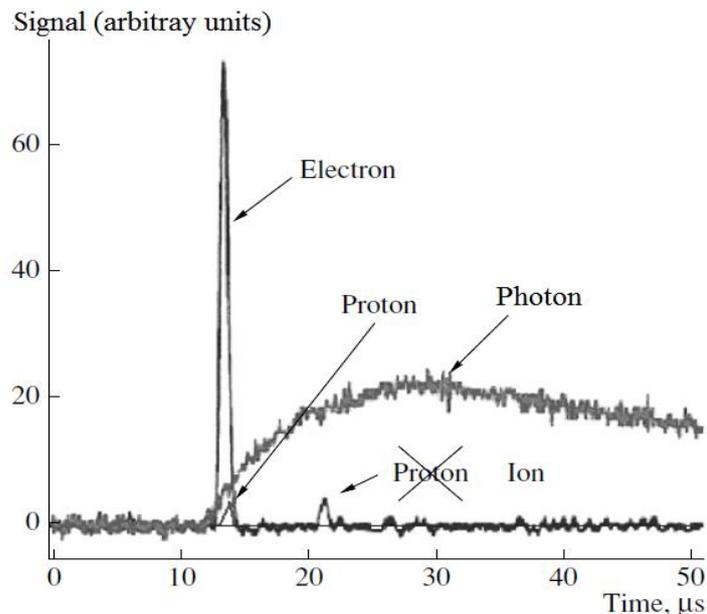

**Fig.5**. The signal from the decay proton has to be delayed by less than one microsecond, which is why it is located at the base of the electron pulse (see line number 2) and so cannot be registered by the combined electron-proton detector. The pulses that are delayed by longer than 1 microsecond are pulses not from decay protons, as it was indicated in ref. [7], but rather from ions, formed in the decay zone. The line number 1 shows the shape of pulses from the gamma detector.

**Analysis and comparison of triple coincidences spectra**

In paper [11] the emiT group researched only the ordinary decay mode, thus this comparison is limited to our spectrum of triple coincidences, presented in Fig.5, and the only peak published by the NIST authors in Nature [7], presented on Fig.7. Analysing the double coincidences spectra obtained by our and the emiT groups (both of which present two main peaks) shows that in the spectrum of triple coincidences we should observe not two but three peaks. Namely, along with the sought after radiative peak, the triple e-p-gamma coincidences spectrum should show two (not one!) response peaks to the registration of beta-electrons and the registration of protons. Fig.5 of triple coincidences clearly shows three peaks, and the leftmost peak with the maximum in channel 103 is connected to the peak of the radiative gamma-quanta in question, as this gamma-quantum is registered by the gamma detectors in our equipment before the electron.

It is also important to note that while both teams' double coincidences spectra show the peaks at a distance from each other and easily distinguishable, in the spectrum of the triple coincidences the radiative peak is on the left slope of the response peak to electron registration. This means that we observe the peak of radioactive neutron decay events against a heterogeneous background. At the same time, both response peaks on the spectrum of triple coincidences are significantly wider and located closer to each other than in the original spectrum of double coincidences. As we demonstrate below, one must take into account such spectrum behavior (related to the presence of a response in the electron detector system of data collection) by introducing the non-local response function. Using this well known method it is possible to distinguish $N_T$ the number of triple coincidences from the heterogeneous background, arriving at the experimental BR value.

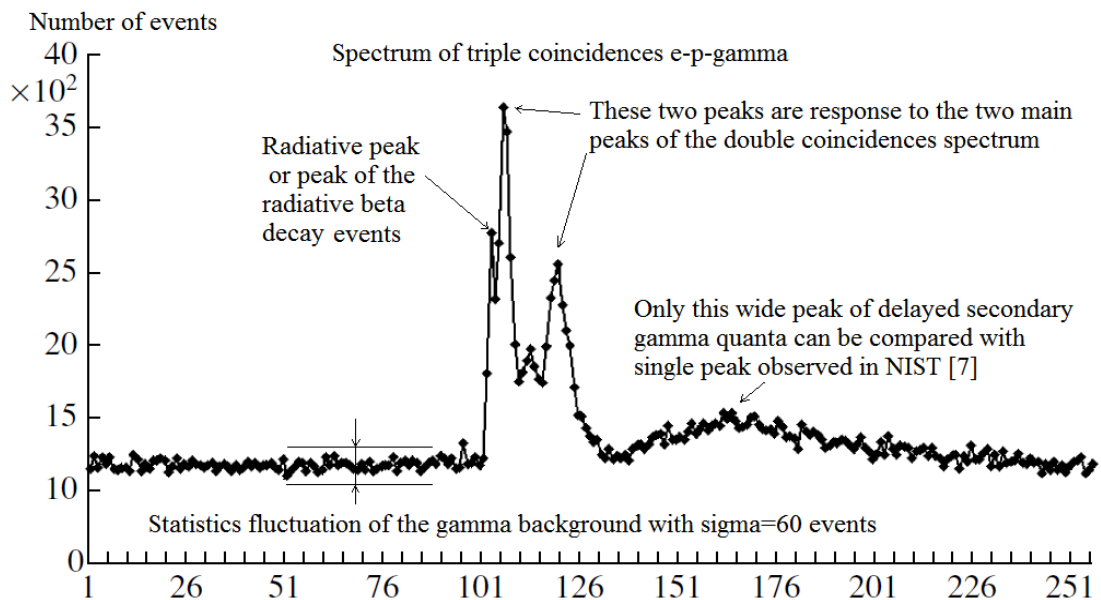

**Fig.6.** Timing spectrum for triple e-p-g coincidences. Each channel corresponds to 25 ns. In this spectrum, three main peaks in channels 103, 106 and 120 can be distinguished. The leftmost peak in 103 channel among these three main peaks is connected with the peak of radiative decay events.

Comparing Fig.2 and 6, it becomes clear that if we ignore the first leftmost peak with the maximum in channel 103 in Fig. 6, the spectrum of double e-p coincidences will resemble the spectrum of triple e-p-γ coincidences on Fig.2. The peak with the maximum in channel 106 on Fig.6 is connected to the left peak of false coincidences on Fig.2, and the peak with the maximum in channel 120 on Fig.6 is connected to the right peak of e-p coincidences on Fig.2. The emerging picture becomes obvious when one uses a standard procedure, introducing a response function for gamma channel $R_\gamma(t,t')$ [6], which is also necessary for calculating the number of triple radiative coincidences $N_T$ in radiative peak:

$$S_{out}(t)=\int S_{in}(t')R_\gamma(t,t')dt'$$

Using the method of response function, one can confidently define our double-humped background: the narrow peak with the maximum in channel 106 on Fig.6 is the response to the narrow peak of zero coincidences (by other words this peak is response to beta-electron registration) in channels 99-100 on Fig.2, and the second peak in this double-humped background on Fig.5 is the response to the peak in channels 117-127 on Fig.2 (or this peak is response to proton registration).

It must be noted that in our case we have to use the non-local response function, as the peaks on the original spectrum $S_{in}(t)$ of double coincidences are significantly narrower than those in the spectrum $S_{out}(t)$ of triple coincidences and also are shifted in their relative positions. In this case we use "functional" multiplication, however if we use the local response function, the triple coincidences spectrum is arrived at by simple multiplying the double coincidences spectrum by a number, in which case neither the width of the peaks nor their position change. It is also evident that the local response function approach leads to an erroneous number of triple coincidences $N_T$ and, therefore, the wrong BR value.

When discussing the similarities between the spectra on Fig.2 and Fig.6, it is important to note that the response peak on Fig.6 with a maximum in channel 106 is shifted to the right or delayed in comparison to the peak responding to electron registration in channel 100 on Fig.1. This is due to the fact that in our electron diagram we used a constant fraction discriminator (CFD). CFD has its own delay line and the location of the time-pickoff signal it generates is determined by the method of comparing the fraction of the original signal to the delayed (CFD method [14]). Thus, there is a shift in the first response peak with a maximum in channel 106 on Fig.6 versus the first peak on Fig.2 with the maximum in channel 100. The value of this delay is equal to the front duration of the gamma quantum signal and is on average 150 ns. The CFD method obviously also shifts the radiative peak, but it should be located to the left of the response peak, as is observed on Fig.6.

As for the wide, almost indistinguishable peak in channel 165 on Fig.6, its influence on the radiative peak in channel 103 is negligible. Its nature is in no way related to the researched phenomenon, so we do not discuss it in our article. This peak is created by the radioactive gamma quanta delayed on average by 1.25 μs and emitted by the radioactive medium within our experimental equipment. The medium is activated by registered beta-electrons. This event of artificial, induced radioactivity has been known for over 100 years and does not have anything in common with the new event of radiative neutron decay which is the subject of current research. As we will demonstrate below, only this 1 microsecond peak and delayed from the registration time by about the same time can be compared to the peak observed by the authors of paper [7] at NIST (see Fig.7). Thus, the authors of this experiment observed not the events of radiative decay but rather the event of artificial radioactivity, already well known in the time of Joliot-Curie.

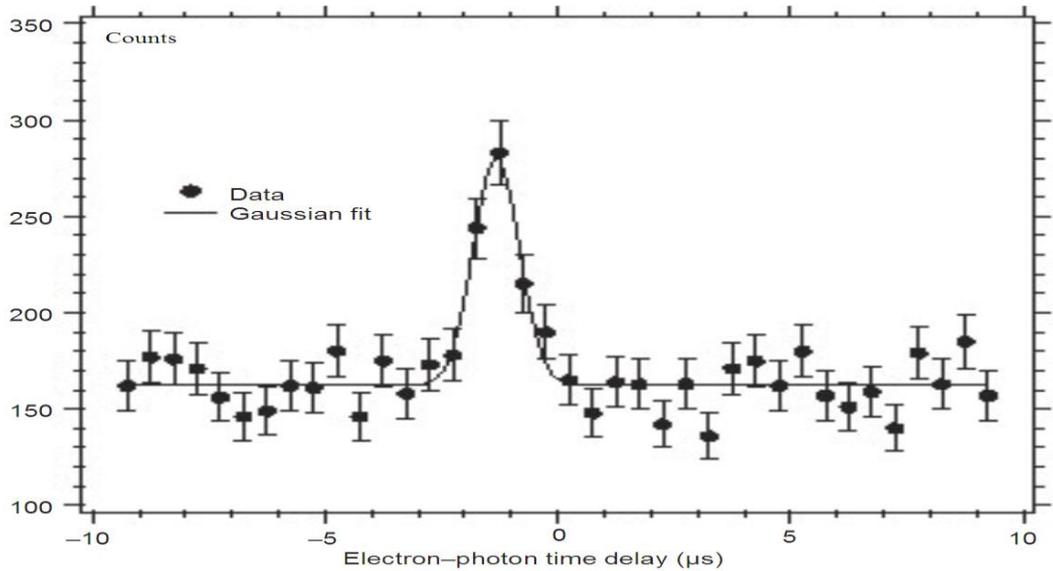

**Fig.7.** Only peak of "electron-photon" coincidences, shifted to the left of 0 – the time of beta-electron registration – by 1.25 microseconds, published in [7, 11].

After analyzing the spectra with the help of the non-local response function $R_\gamma(t,t')$ we finalize the average value for the number of radiative neutron decays $N_T$=360 with a statistics fluctuation of 60 events. B.R. can be expressed as a ratio of $N_T$ to $N_D$ as BR = k ($N_T / N_D$), where coefficient k=3.3 is the geometrical factor that we can calculate by using anisotropic emission of radiative gamma-quanta [4]. With the number of observed double e-p coincidences $N_D = 3.75 \cdot 10^5$ and triple e-p-γ coincidences $N_T = 360$, one can deduce the value for radiative decay branching ratio of $(3.2 \pm 1.6) \cdot 10^{-3}$ (99.7 % C.L.) with the threshold gamma energy ω=35 keV. The average B.R. value we obtained deviates from the standard model, but because of the presence of a significant error (50%) we cannot make any definite conclusions. The measurements must be made with greater precision. According to our estimates, in the future experiment we will be able to make more definite conclusions about deviation from the standard electroweak theory with experimental error less than 10%.

The difference between the NIST experiment and our experiment becomes immediately apparent. First and foremost, it is the time scale: in our spectra, the scale is measured in nanoseconds, while in the other experiment the scale is in microseconds. Besides, we used three types of detectors, each of which registered its own particle: one detector for the electrons, one for the protons, and six identical detectors for the radiative gamma-quanta (see [6]). The duration of the front pulse from the electron and proton detectors is 10 nanoseconds in our experiment and 100 times greater than that in the NIST experiment, in the order of 1 μs. The rise time of gamma signal from our gamma-detectors is on average 150 ns, and from avalanche diode on the NIST equipment greater than 10 μs, besides that the diode pulse arrives with significant noise, which makes the thickness of the front pulse line equal to more than 0.5 μs (see the photon line on Fig.7 from [7]). All of this leads to our factual time resolution being two orders of magnitude better than the resolution achieved in the NIST experiment. However, as the two experiments used equipment which was practically the same in size and smaller than 1 meter, the choice of the time scale is a matter of principle. Given this geometry, it is impossible to get microsecond signal delays from all of the registered charged particles, i.e. electrons and protons. In this light, it is surprising that the peak identified by the authors of the NIST report [7] as the peak of radiative gamma-quanta, is shifted by 1.25 microseconds to the left. The expectation that magnetic fields of several tesla in magnitude delay all electrons and protons, are absolutely ungrounded.

Indeed, the magnetic field cannot change the speed of charged particles. It can only twist a line trajectory into a spiral. The length l of this spiral depends on angle θ between particle velocity and magnetic field direction. In beta decay, electrons can fly out under any angle θ, therefore the magnetic field can increase the time of delay by several orders of magnitude only for a negligible portion of the charged particles. Even this negligible number of particles that flew out at an almost 90 degree angle to the direction of the magnetic field that coincides with the direction of the narrow neutron guide (see Fig.4) will most likely end up on the walls of the neutron guide rather than reach and hit the detector due to the presence of the strong electrostatic field. Because the distance between the point of decay and the detector is about 0.5 meter and electron velocity is comparable with speed of light, the electron time of delay should be less than a microsecond by two orders of magnitude.

Thus, both the 1 microsecond shift and the width of the only peak on Fig.7 in the experiment conducted at NIST, is in sharp contradiction to elementary estimates. We, on the other hand, did not observe any wide peaks before electron registration and our gamma background is very even in this part of the spectrum (see Fig.6). However, when we assume that the NIST experiment authors observed the wide peak, shifted by 1 microsecond, not before, but after the registration of beta-electrons. In that case, the wide peak on our spectrum in Fig.6 completely corresponds to the wide peak on Fig.7. However, as noted above, density of gas molecules remains high even with the pressure of $10^{-8}$- $10^{-9}$ mbar and this residual gas is activated by beta-electrons. The wide peak in our spectrum is formed by the delayed gamma quanta from this induced artificial radioactivity.

**Conclusions.** The main result of our experiment is the discovery of the radiative peak namely in the location and of the width that we expected. The location and the width of the radiative peak correspond to both estimates and the detailed Monte Carlo simulation of the experiment. Thus, we can identify the events of radiative neutron decay and measure its relative intensity, which was found to be equal B.R. = $(3.2\pm1.6)\cdot 10^{-3}$ (with C.L.=99.7% and gamma quanta energy over 35 keV ).

At the same time, the average experimental B.R. value exceeds the theoretical value by 1.5 times. However, due to a significant error we cannot use this result to assert that we observe a deviation from the standard model. Therefore, our most immediate goal is to increase experiment precision, which we can improve by several percents according to estimates.

For last two years we were preparing this new experiment and conducted number of tests for our new electronics. We constructed multi channel generator what can generate the pulses with the same forms as our electron, proton and gamma detectors. During these tests we got the same responses as during our last experiment on real neutron beams at FRMII. It means that all additional peaks on our spectra have no any physics reasons and It proves once more that we were absolute correct when applied the response function method for explaining these peaks as response ones and for developing our experimental spectra.

We created and tested our new electronic system for obtaining experimental spectra. By using this new programmable electronics we can significantly reduce the influence of response peaks on peak with radiative decay events. Now we can get this peak almost isolated from responses. On our estimations all these allow us to reach accuracy for our new experiment about 1%. So, on the base of our new electronics we can confirm or refuse the deviation of our average experimental value of BR from the standard model one.

As concerning the comparison of our experimental results with others we can make the following two main conclusions. The main parameters of our spectrum of double electron-proton coincidences identifying the events of ordinary neutron decay fully coincide with an analogous spectrum published by emiT group in [11].

Unfortunately we cannot say same for another experiment measuring the radiative neutron decay published in [7]. Particularly vexing is the authors' unsubstantiated assertion that they observe their only wide peak of gamma quanta before the registration of beta-electrons. Both the position and the width of this peak are located in sharp contradiction to both the elementary estimates, and the results of our experiment. In the course of our entire experiment we did not observe such a wide peak in the triple coincidences spectrum, located before the arrival of electrons at a huge distance of 1.25 µs. However, it is possible to reconcile our spectra of triple coincidences with the one isolated peak observed at NIST if we assume that at NIST, the gamma-quanta were registered after the beta electrons. Only in this case does the NIST peak almost completely coincide with the peak we observed in the spectra of triple coincidences with the maximum in channel 165, both in terms of the huge delay of 1.25 µs and in terms of its huge width. This peak is created by the delayed secondary radioactive gamma-quanta, arising from the activation by beta electrons of the media inside experimental chamber, which was the real object of the NIST experimentalists' observation.

Despite the recent disagreements [15], which we consider to be subjective in nature [16], we acknowledge the contribution of our Western colleagues Profs. N. Severijns, O. Zimmer and Drs. H.-F. Wirth, D. Rich to our experiment conducted in 2005. Here it is important to note that the authors of the article published in Nature [7] consciously misled first our Western colleagues and then the physics community at large by insisting that their only wide peak is removed by 1.25 microseconds to the left from the time of electron registration, when in reality this peak was formed by delayed gamma-quanta, emitted by the activated air inside the experimental equipment, and corresponds to our wide peak with the maximum in channel 165 (refer to Fig.6) [15, 16]**.** The authors would like to thank Profs. D. Dubbers and Drs. T. Soldner, G. Petzoldt and S. Mironov for valuable remarks and discussions. We are also grateful to the administration of the FRMII, especially Profs. K. Schreckenbach and W. Petry for organizing our work. We would especially like to thank Honorary President of NRC "Kurchatov Institute" Academician E.P. Velikhov (see Appendix), Academishian S.S. Gershtein and Prof. P. Depommier, V.P. Martem'yanov for their support, without which we would not have been able to conduct this experiment. Financial support for this work was obtained from RFBR (Project N 07-02-00170).

APPENDIX
This article was initially sent to Nature Physics and later on to Nature journals. The related correspondence with journals' editorial boards and a support letter from Honorary President of NRC "Kurchatov Institute" follow below.

Rashid Khafizov
Moscow, Russian Federation
______________________________________________________________________________
__________

Editorial Board
NATURE
International journal of science

Date: April 2018

Dear Editorial Board members,
In this letter, I am referring to the radiative neutron decay. When you have a spare minute, please google the above key reference words. In few seconds, the search engine will produce a hit list topped by the links to the documents describing our experiment conducted in 2005 at the FRM II facility operated by the Technical University of Munich.

This experiment discovered the radiative neutron decay. For the first time, we were observing its events by the triple coincidences of the recoil proton, electron, and radiative gamma-quantum.

In 2006, the experiment was repeated by the NIST group led by Mr James Byrne, the only group that re-ran this experiment after us. However, both the 2006 experiment and their latest experiment conducted in 2016 did fail to identify both the events of the radiative neutron decay and the events of the ordinary beta decay.

Although the researchers bluntly manipulated the results of their experiment, the NATURE Letters Editorial Board found it appropriate to publish the very first work by Mr James Byrne et al. asserting to have identified the radiative neutron decay. In actual fact, the group were merely observing the process of ionization taking place in the magnetic field (see Nature Letters, vol. 444, 21/28 December 2006, page 1059, and Attachment 1 – our submitted article herewith).

The points outlined below explain why we believe the results of the Byrne group experiment cannot be recognized as valid and what process was actually observed during the NIST experimental group (emiT group) experiment.

1.      The paper's authors cannot even register the events of the ORDINARY neutron decay by coincidences of a beta electron with a beta decay proton. Instead of the double coincidences of an electron with beta decay proton, the authors register coincidences of electrons and ions. The number of protons emitted during neutron beta decay is negligible as compared to the number of ions produced by the strong ionizing radiation when the intense beam of cold neutrons transits through the experimental chamber. More so, the experiment group applied strong magnetic fields that mixed a small number of protons emitted by beta decay with a huge number of ions.

Consequently, the experimentalists were not able to reproduce either our result or the result obtained by the emiT group. This can be clearly seen in the double coincidences spectra (see Attachment 2). Fig. 2 shows our spectrum, Fig. 3 shows the emiT group spectrum. The emiT group experiment conducted at NIST also registered double coincidences of a beta electron with beta decay proton; however, it was not aimed to search for radiative neutron decay by triple coincidences. The coincidence of our spectrum with that of the emiT group is self-explanatory and needs no saying. The horizontal level of the ion background demonstrates two peaks - a response peak occurring in the aftermath of the electron registration and a peak brought about by the delayed protons emitted by neutron beta decay, the number of neutrons being the number of detectable ordinary neutron beta decays. The spectrum obtained by the Byrne group (see Fig. 4) and our spectrum are different as chalk and cheese. The huge Byrne group peak is, beyond any doubt, formed by ions, and the lightest ions – protons – are not traced at all while the dead time is significant. Therefore, no registration is ongoing at the beginning of the spectrum.

2. Significant dead times are a consequence of the use of the combined electron-proton detector. An electron has huge energy and it generates a powerful signal, but the detector registers nothing at all throughout the whole signal transition time! (See Attachment 2, Fig. 5.)

3. While registering ions instead of protons emitted by beta decay, the experimentalists are overlooking the peak produced by the radiative neutron decay in the triple coincidence spectrum which represents a much more intriguing and subtle feature. They experimentalists use slow avalanche diodes (see Attachment 1, Fig. 2, where the photon impulse front duration is 15 microseconds!). Hence they have a timescale expressed in microseconds, and not in nanoseconds. In the case of the radiative decay, the delay times are tens of nanoseconds. That is why we observed a radiation peak on the left slope of the peak-response occurring in the aftermath of the electron registration as the electron and gamma-quantum enter the detectors virtually simultaneously. We did not observe any wide peak with a width of 1 microsecond ahead of the registration of the electron (see the spectrum in Figure 2, Attachment 2, where all peaks are marked). Only downstream of the electron registration did we observe a wide peak formed by gamma-quanta occurring in the aftermath of the rarefied air ionization inside the reactor. It is to be noted that the narrow radiation peak is separated in time from the broad peak of the artificial gamma radioactivity, and the process associated with the emission of gamma-quanta during ionization is delayed in comparison with the process of neutron radiative decay.

4. Therefore, the process registered by the Byrne group has nothing to do whatsoever with the radiative neutron decay. It rather relates to the radiation of gamma-quanta during the rarefied air ionization. This phenomenon can be observed in the form of polar lights occurring in the rarefied atmosphere in the Earth's geomagnetic field.

5.     The ionization process is also a noteworthy process. Our experiment and the Byrne group experiment demonstrate that this process can and should be investigated using an intense beam of cold neutrons. But the Byrne group bluntly manipulated the results of the experiment by manually placing their peak of triple coincidences ahead of the moment of electron registration and describing it as a radiation peak, thus creating a painful deadlock impeding the scientific process. In the years that followed, their intrigues, manipulations, rows, letters to the editors and phone-calls to the members of the Academy of Sciences turned things even worse. The Byrne group's latest work published in 2006 contains nothing beyond what we already know. We see the same experimental equipment with the same strong magnetic fields, dead times and you name it. The word "ion" is nowhere to be found in the works by the Byrne group which turns a blind eye to the process of ionization and, specifically, the ion background.

In view of the foregoing, may I kindly ask the Editorial Board to consider publishing our work which I attach herewith. This publication will help us greatly find a way out of the current impasse where our work which tops the hit list of the Google search engine is being officially ignored. As we see it, there are two ways to go forward – to pursue a study of the radiative neutron decay and a study of the ionization process. The publication of our work in Nature Letters will make a valuable contribution to the promotion of our new radiative neutron decay experiment. In fact, our new experimental equipment is operational, and we have to get it to the beam. We are open to various options and will be willing to conduct a new experiment using our new hardware in ILL, FRM II (Munich) or even NIST provided we receive international support.

With my best regards,
Rashid Khafizov

Dear Dr Andreas Trabesinger,

Ref. NPHYS - 2018 - 05 - 1460
Attached for your attention is a letter by Academician E.P. Velikhov, Honorary President of the Kurchatov Institute National Research Centre. It will be very much appreciated if you could read the letter and kindly bring its content to the attention of the Editorial Board. If required, the original will be forwarded to your with currier mail.
With my best regards,
Rashid Khafizov

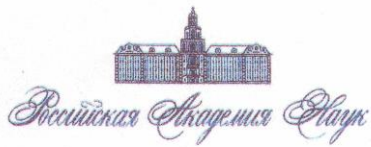

Академик

Евгений Павлович
Велихов

Dear Nature Group editors,

First of all, allow me to explain briefly why I am paying scientific interest and giving my support to a publication of the paper "The time spectra in the radiative neutron decay experiment" prepared by the researchers of our Centre.

1. The event of the radiative neutron decay by triple coincidences of the recoil proton, electron, and radiative gamma-quantum was first registered by the authors of this paper. Not only did they measure the relative intensity (BR), which represents the main feature of the radiative neutron decay, but they also recorded events of the ordinary neutron decay by double coincidences of the electron and recoil proton. Division of the number of triple coincidences by the number of double coincidences produces the BR experimental value. In the meantime, the authors of another paper, first published in Nature (J. Byrne, et.al. Nature Letters, vol. 444, 21/28 December 2006, page 1059), could not even observe the event of the usual neutron decay, because instead of the protons the ions were recorded. More so, while failing to observe the radiative decay of the neutron, they registered the ionization process with gamma-ray emission.

2. Further, not only did the authors of the paper which I am supporting measure the BR but they experimentally observed a DEVIATION FROM THE STANDARD MODEL, specifically a deviation of their experimentally measured BR value from the BP theoretical value computed by Yu. V. Gaponov and R. U. Khafizov, researcher of our Centre, back in 1994. Subsequently, these BR theoretical values were confirmed by other researchers. However, the mean BR experimental value gained by our experiment group is significantly greater than the standard model theoretical value. In contrast, the J. Byrne group, while manipulating their measurements by the ionization of residual air in their experimental chamber, are claiming their BR experimental value agrees with the standard model and the theoretical value first calculated by Yu.V. Gaponov and R.U. Khafizov.

3. Considering the deviation from the standard model recorded by our group and recognising that this deviation is greater than that in the standard model, it is reasonable to believe that there are "spare" radiation gamma-quanta present. This brings up the question: Why and wherefrom do the "spare" gamma-quanta appear? This question can be answered as follows: They are generated directly inside the decaying neutron. These are structural gamma-quanta and their characteristics can give information about the very structure of the elementary particle which is of paramount interest and which makes the study of the neutron's radiation nature a most intriguing objective in modern Physics.

To sum up the above, I would like to recall the fundamental principle which ensures the objective development of science as a whole. The point of view described in the paper by our group represents an alternative to the point of view expressed by Mr J. Byrne et al. in their paper published in Nature. Therefore, an alternative view has to be offered to the scientific community, preferably in Nature or other Nature Group publications.

Otherwise the Editorial Board will have to assume responsibility for favouring one party which, in fact, has studied a completely different process, namely the ionization process instead of the neutron radiative decay. The publication of the paper expressing our researchers' point of view is needed right now. Quite recently, Mr J. Byrne and his group had run the second experiment where they had made the same blunt errors that were made during their original experiment described in the paper published by Nature 10 years ago.

With my best regards,

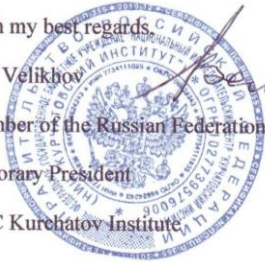

E.P. Velikhov

Member of the Russian Federation Academy of Sciences

Honorary President

RNC Kurchatov Institute

Dear Dr Andrea Taroni, dear Dr Andreas Trabesinger,

I still would like to respond to the two points raised in your last letter.

Firstly, we sent our article to you because, on the one hand, you are a part of the Nature group of journals and, on the other hand, you are not a general journal, your specific interest area lies in Physics which is reflected in the title of your journal. This is the reason why we wanted to draw your attention to the fact that the article published in Nature by Mr Byrne et aliae contains gross mistakes relating to the recording of an entirely different effect. They study the effect of gamma-ray emission in the ionization process; this effect underlies the well-known "polar aurora" phenomenon. In Russia, even school students are taught how the density of the atmosphere varies depending on the altitude (see https://www.slideserve.com/torgny/669, specifically slide 14); what causes the appearance of the polar aurora, at what altitude it occurs. So, the density of the residual atmosphere, which was in our experimental chambers, corresponds to a height of over 150 km above the ground, where the aurora is observed. Mr Byrne et alias observed merely the phenomenon of the aurora (or, as it is indicated in Prof. Velikhov's letter "gamma-ray emission during the ionization of the residual gas in the chamber"), which they advertise with the help of Nature as a completely different UNRESEARCHED phenomenon - the radiative neutron decay. They have all the conditions for studying the Aurora, there is a magnetic field, a rarefied atmosphere of appropriate density and ionizing radiation. But the authors of the article in Nature do not mention the well-known ionization process; you will not even find the word ion in this article and their subsequent articles. Why should we reduce everything to such an insignificant, I would say, anecdotal, level? The foregoing clearly demonstrates that the authors of the article published in the Nature has created a scandalous, deadlock situation; we need no scandal, we need support in the form of our work publication. This will be a way out of the created impasse which will enable us to run a new experiment.

Secondly, we are not insisting on the word review. Please, you can publish our work as an article or a critical article. As indicated in the letter of Prof. Velikhov, we do represent the alternative point of view versus the one published in Nature. We would like to note that the radiative decay and polar aurora observed by Mr Byrne et alias are absolutely different phenomena. How can they possibly be "balanced"?! Why our work is of interest for "the specific circle of your readers" is clearly explained in the letter of Prof. Velikhov too. For our part, we took into account the fact that your journal is aimed precisely at the WIDE circle of readers interested in Physics. In our article, you will not find any "mathematical tediousness," no kilometer long formulae. On the contrary, we, using the charts, explain in detail what the response function is all about, for example, the difference of the functional multiplication and the simple multiplication of a function by a number. I assure you that throughout the world a huge number of school students reading your journal understand what the functional multiplication is all about; they are familiar with the word ion and with the process of ionization, that this process underlies a most beautiful phenomenon of the polar aurora.

Summarizing the above let me assure you we will not let our work be silenced. If nevertheless our article is not published, the broad scientific community will remain incognizant of our work. We will be forced to publish it in the Internet explaining in the annex thereto where we sent the article to and what arguments were brought in correspondence with you. We do not want this outcome. The fact is that the equipment for a new neutron radiative decay experiment is ready; all we need is to get to the intense cold neutrons beam. For this purpose, we just need the publication in your journal. And at the informal level, our work is known; google keywords "radiative neutron decay" and it will top of the hit list.

Regardless of our arguments both Nature and Nature Physics declined to send the article for peer review.